\documentclass[twocolumn,showpacs,preprintnumbers,amsmath,amssymb]{revtex4}
\usepackage {longtable}
\usepackage{graphicx}
\usepackage{dcolumn}
\usepackage{bm}

\begin{document}

\title{Ionization of Sodium and Rubidium \textit{n}S, \textit{n}P and \textit{n}D Rydberg atoms by blackbody radiation}

\author{I.~I.~Beterov}
 \email{beterov@isp.nsc.ru}
\author{D.~B.~Tretyakov}
\author{I.~I.~Ryabtsev}
\affiliation{Institute of Semiconductor Physics, Pr.Lavrentyeva
13, 630090 Novosibirsk, Russia }

\author{A.~Ekers}
\affiliation{University of Latvia, Institute of Atomic Physics and
Spectroscopy, LV-1586 Riga, Latvia }
\author{N.~N.~Bezuglov}
\affiliation{ St. Petersburg State University, Fock Institute of
Physics, 198904 St.-Petersburg, Russia }
\date{February 20, 2007}

\begin{abstract}
Results of theoretical calculations of ionization rates of Rb and
Na Rydberg atoms by blackbody radiation (BBR) are presented.
Calculations have been performed for \textit{n}S, \textit{n}P and
\textit{n}D states of Na and Rb, which are commonly used in a
variety of experiments, at principal quantum numbers
\textit{n}=8-65 and at three ambient temperatures of 77, 300 and
600~K. A peculiarity of our calculations is that we take into
account the contributions of BBR-induced redistribution of
population between Rydberg states prior to photoionization and
field ionization by extraction electric field pulses. The obtained
results show that these phenomena affect both the magnitude of
measured ionization rates and shapes of their dependences on
\textit{n}. The calculated ionization rates are compared with the
results of our earlier measurements of BBR-induced ionization
rates of Na \textit{n}S and \textit{n}D Rydberg states with
\textit{n}=8-20 at 300~K. A good agreement for all states except
\textit{n}S with $n>15$ is observed. We also present the useful
analytical formulae for quick estimation of BBR ionization rates
of Rydberg atoms.
\end{abstract}

\maketitle




\section{Introduction.}

Blackbody radiation (BBR) is known to strongly affect the
populations of atoms in highly excited Rydberg
states~\cite{Gallagher}. It has also been shown that at the
ambient temperature of 300~K BBR can photoionize Rydberg atoms
with $n \sim 20$ at astonishingly high rates ($\sim
10^3\,\mathrm{s}^{-1}$)~\cite{SpencerIon, Beiting, Spencer}.
Strong effect of BBR on Rydberg atoms is related to large matrix
elements of bound-bound and bound-free transitions between Rydberg
states in the microwave and far infrared spectral
range~\cite{Rydberg atoms}.

Interaction of Rydberg atoms with BBR has been studied earlier in
various contexts. Farley and Wing~\cite{Farley} calculated the
dynamic Stark shifts and depopulation rates of Rydberg levels of
alkali atoms with $ n\le 30$ at 300~K. The temperature dependence
of BBR-induced transitions rates from the 19S state of sodium was
calculated and measured by Spencer et al~\cite{Spencer}. Galvez et
al~\cite{Galvez1, Galvez2} have studied BBR-induced cascade
transitions from the initially populated $n$=24-29 states of Na,
both theoretically and experimentally.

Although interaction of Rydberg atoms with blackbody radiation has
been studied for years, both theoretically and experimentally,
only a few works were devoted to BBR-induced ionization of Rydberg
atoms itself. The temperature dependence of the BBR-induced
ionization rate of Na~17D state was numerically calculated using
the quantum defect method in a Coulomb approximation and
experimentally measured by Spencer et al~\cite{SpencerIon}. A
simple scaling law for BBR ionization rates was also introduced in
that work. More recently, the interest to BBR-induced ionization
of Rydberg atoms has been related to the spontaneous formation of
ultracold plasma in dense samples of cold Rydberg atoms
\cite{Robinson, Li}, and to the prospects of its use as a
convenient reference signal in absolute measurements of
collisional ionization rates \cite{PaperI}. Even nowadays,
however, the studies of spontaneous evolution of ultracold Rydberg
atoms to a plasma caused by BBR \cite{Robinson,Li} use the simple
estimates of BBR ionization rates taken from the well-known work
\cite{SpencerIon}. The numerical calculations of direct BBR
photoionization rates were performed by Lehman \cite{Lehman} for H
and Na Rydberg states with $n$=10-40 using a Herman-Skillman
potential with an additional term to account for core
polarization. However, the theoretical data for Rb atoms, which
are widely investigated in experiments on ultracold plasma, are
lacking. No systematic experimental studies of the $n$-dependences
of BBR ionization rates of alkali-metal Rydberg atoms in a wide
range of principal quantum numbers are known.

A major problem in the interpretation of measured ionization rates
of Rydberg atoms is related to the fact that populations of
Rydberg states are affected by BBR-induced processes to an unknown
(or, at least, not straightforwardly predictable) extent, which
depends on the combination of specific experimental conditions
(principal quantum numbers \textit{n}, ambient temperature,
duration of measurements, extracting electric field strengths,
etc.). Therefore, a more detailed study of BBR-induced ionization
of Rydberg atoms under typical experimental conditions is
required, especially for higher Rydberg states that are often
explored in the experiments with cold atoms.

In this article we present the results of numerical calculations
of BBR ionization rates of Rb and Na Rydberg atoms for the most
commonly used \textit{n}S, \textit{n}P and \textit{n}D states with
\textit{n}=8-65 at the ambient temperatures of 77, 300 and 600~K.
In these calculations we take into account two phenomena that may
affect the observed ionization rates: the time-dependent
BBR-induced population redistribution between Rydberg states prior
to photoionization and the selective field ionization (SFI) of
high-lying Rydberg states by the electric field applied for
extraction of ions from the excitation zone. Although both
phenomena were mentioned in some of the earlier
studies~\cite{SpencerIon}, their effects have not yet been studied
in sufficient detail neither theoretically, nor experimentally. In
particular,  in the measurements of direct BBR photoionization
rate of the sodium 17D state \cite{SpencerIon} with effective
lifetime 4~$\mu$s  at the ambient temperature of 300~K, the effect
of population redistribution was  diminished due to the short
measurement time of 500~ns and  weak electric field used for
extraction of ions. However, for Rydberg states with $n \sim 8 $
and effective lifetimes less than 500~ns even this very short time
interval is insufficient to avoid mixing with neighboring states,
so that mixing processes must be necessarily accounted in
calculations. In the present study we address to both phenomena
and provide the useful analytical formulae, which can be applied
to any Rydberg atom for estimates of the direct BBR-induced
ionization rates and of the contribution of the SFI.

The simplest (but often insufficient) way of considering BBR-induced
ionization after the excitation of an atom \textit{A} to a given \textit{nL}
Rydberg state is to limit the problem to the direct photoionization of the
initial \textit{nL} state in one step by absorption of BBR photons:

\begin{equation}
\label{eq1}
A\left( {nL} \right) + \hbar \omega _{BBR} \to A^{ +}  + e^{ -} ,
\end{equation}

\noindent where $\hbar \omega _{BBR} $ is the energy of absorbed
BBR photon, $A^{+}$ is the atomic ion, and $e^-$ is the free
electron emitted in the ionization. In the reality, however,
ionization of Rydberg atoms exposed to BBR is a complex process,
in which the following main components can be identified [see
Fig.~1(a)]: (i) direct photoionization of atoms from the initial
Rydberg state via absorption of BBR photons, (ii) field ionization
by extraction electric field pulses of high Rydberg states, which
are populated from the initial Rydberg state by absorption of BBR
photons, (iii) direct BBR-induced photoionization of atoms in the
neighboring Rydberg states, which are populated due to absorption
and emission of BBR photons prior to photoionization, and (iv)
field ionization of other high-lying states, which are populated
via population redistribution involving two or more steps of BBR
photon absorption and/or emission events. Our calculations show
that all these processes can contribute to the total ionization
rate to a comparable extent, and, therefore, none of them can be
disregarded. In what follows we will consider the above processes
separately and calculate the total BBR ionization rates, both
analytically and numerically.




\section{Calculation of BBR ionization rates.}

Ionization mechanisms of Rydberg atoms exposed to BBR are
illustrated in Fig.~1. The total BRR-induced ionization rate can
be written as consisting of four separable contributions:

\begin{equation}
\label{eq2}
W_{BBR}^{tot} = W_{BBR} + W_{SFI} + W_{BBR}^{mix} + W_{SFI}^{mix} .
\end{equation}

\begin{figure}
\label{Fig1}
\includegraphics[width=7cm]{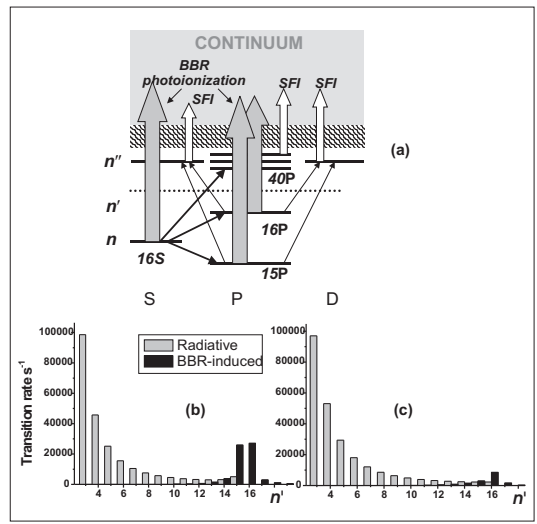}
\caption{(a) Schematic illustration of BBR induced and field
ionization processes occurred after excitation of the initial
Na(16S) state, including redistribution of population over other
$n'L'$ Rydberg states due to spontaneous and BBR induced
transitions from the initial state. (b) Calculated spontaneous and
BBR induced transition rates from the initial 16S state to other
$n'$P states. (c) Calculated spontaneous and BBR induced
transition rates from the initial 16D state to other $n'$P
states.}

\end{figure}


The first contribution, $W_{BBR} $, is the direct BBR
photoionization rate of the initially excited \textit{nL} state,
which  will be discussed in Section~II~A. The second term,
$W_{SFI} $, is the rate of the selective field ionization (SFI) of
high $n''L'$ Rydberg states, which are populated from the initial
Rydberg state \textit{nL} via absorption of BBR photons. This
field ionization will be discussed in Section~II~C, while
redistribution of population between Rydberg states will be
described in Section~II~B. The third term, $W_{BBR}^{mix} $, is
the total rate of BBR-induced photoionization of neighboring
$n'L'$ Rydberg states, which are populated via spontaneous and
BBR-induced transitions from the initial state. The last term,
$W_{SFI}^{mix}$, is the rate of SFI of high-lying Rydberg $n''L'$
states that are populated in a two-step process via absorption of
BBR photons by atoms in $n'L'$ states (note, that here, in
contrast to $W_{SFI} $, we consider lower $n'L'$, states which
cannot be directly field ionized). These latter two ionization
rates, which are related to population redistribution between
Rydberg states, will be considered in Section~II~D . The atomic
units are used below, unless specified otherwise.

\subsection{Direct BBR photoionization.}

Direct BBR-induced photoionization rate $W_{BBR} $ of a given
\textit{nL} state is calculated from the general formula
\cite{SpencerIon}:

\begin{equation}
\label{eq3} W_{BBR} = c\int\limits_{\omega _{nL}} ^{\infty}
{\sigma _{\omega}  \;\rho _{\omega}  d\omega},
\end{equation}

\noindent where \textit{c} is the speed of light, $\omega _{nL} =
1/\left( {2n_{eff}^{2}}  \right)$ is the photoionization threshold
frequency for the \textit{nL} Rydberg state with the effective
principal quantum number $n_{eff}=\left({n-\delta_{L}}\right)$,
where $\delta_{L}$ is a quantum defect, and $\sigma_{\omega}$ is
the photoionization cross-section at the frequency $\omega$. The
volume density $\rho_{\omega}$ of BBR photons at the temperature
\textit{T} is given by the Plank distribution:

\begin{equation}
\label{eq4} \rho _{\omega}  = \frac{{\omega ^{2}}}{{\pi
^{2}c^{3}\left[ {\mathrm{e}^{\omega /\left( {kT} \right)} - 1}
\right]}},
\end{equation}

\noindent where \textit{kT} is thermal energy in atomic units. For
isotropic and non-polarized thermal radiation field, the value of
$\sigma_{\omega}$ is determined by the radial matrix elements
$R\left({nL \to E,L \pm 1} \right)$ of dipole transitions from
discrete \textit{nL} Rydberg states to the continuum states with
\textit{L$ \pm $}1 and photoelectron energy \textit{E}:

\begin{equation}
\label{eq5}
\sigma _{\omega}  = \frac{{4\pi ^{2}\omega} }{{3c\left( {2L + 1}
\right)}}\sum\limits_{L' = L \pm 1} {L_{max} R^{2}\left( {nL \to E,L \pm 1}
\right)} ,
\end{equation}

\noindent where \textit{L}$_{max}$ is the largest of $L$ and $L'$.

The main problem in the calculation of \textit{W}$_{BBR}$ for an
arbitrary Rydberg state is thus to find the values of $R\left( {nL
\to E,L \pm 1} \right)$ and their frequency dependence. In order
to achieve high accuracy of the matrix elements, numerical
calculations should be used. In this work we used the
semi-classical formulae derived by Dyachkov and Pankratov
\cite{Dyachkov}. In comparison with other semi-classical methods
\cite{GDK, Davydkin}, this method is advantageous as it gives
orthogonal and normalized continuum wavefunctions, which allow for
the calculation of photoionization cross-sections with high
accuracy. We have verified that photoionization cross-sections of
the lower sodium S states calculated using the approach of
\cite{Dyachkov} are in good agreement with the sophisticated
quantum-mechanical calculations by Aymar \cite{Aymar}.

Approximate analytical expressions for \textit{W}$_{BBR}$ would
also be useful, since they illustrate how the ionization rate
depends on parameters \textit{n}, \textit{L}, and \textit{T}. Such
expressions can be obtained using the analytical formulae for
bound-bound and bound-free matrix elements deduced by Goreslavsky,
Delone and Krainov (GDK) \cite{GDK} in the quasiclassical
approximation. For the direct BBR-induced photoionization of an
\textit{nL} Rydberg state the cross-section is given by:

\begin{eqnarray}
\label{eq6} \sigma _{\omega}  \left( {nL \to E,L \pm 1} \right) =
\qquad \qquad \qquad \qquad \qquad  && \nonumber \\
 =\frac{{4L^{4}}}{{9cn^{3}\omega} } \left[ {K_{2/3} ^{2}\left(
{\frac{{\omega L^{3}}}{{3}}} \right)  + K_{1/3} ^{2}\left(
{\frac{{\omega L^{3}}}{{3}}} \right)} \right],&&
\end{eqnarray}

\noindent where $K_{\nu}  \left( {x} \right)$ is the modified
Bessel function of the second kind. This formula was initially
derived to describe the photoionization of hydrogen atom. It was
assumed that the formula can be extended to alkali atoms simply by
replacing \textit{n} by $n_{eff} = \left( {n - \delta _{L}}
\right)$, where $\delta _{L}$ is the quantum defect of the Rydberg
state. In the reality, however, its accuracy in absolute values is
acceptable only for truly hydrogen-like states with small quantum
defects. A disadvantage of the GDK model is that it disregards
non-hydrogenic phase factors in the overlap integrals of dipole
matrix elements.

The main contribution to \textit{W}$_{BBR}$ in Eq.~(\ref{eq3}) is
due to BBR frequencies near the ionization threshold frequency
$\omega _{nL} $, because the bound-free dipole moments rapidly
decrease with increasing $\omega $. For Rydberg states with
\textit{n$ \gg $}1 and low \textit{L} one has ($\omega
$\textit{L}$^{3}$/3)$ \ll $1. In this case Eq.~(\ref{eq6}) can be
simplified to the form:

\begin{eqnarray}
\label{eq7} \sigma _{\omega}\left( {nL \to E,L \pm 1}
\right)\approx \qquad \qquad \qquad \qquad \qquad&& \nonumber \\
\approx \frac{{1}}{{9cn^{3}}}\left[{\frac{{6^{4/3}\Gamma
^{2}\left( {2/3} \right)}}{{\omega ^{7/3}}}+\frac{{6^{2/3}\Gamma
^{2}\left( {1/3} \right)}}{{\omega ^{5/3}}}L^{2}} \right].& &
\end{eqnarray}

The combination of Eqs.~(\ref{eq3}), (\ref{eq4}) and (\ref{eq7})
yields:

\begin{eqnarray}
\label{eq8} W_{BBR} \approx \frac{{1}}{{\pi
^{2}c^{3}n^{3}}}\int\limits_{\omega _{nL} }^{\infty}  {\left[
{2.22\,\omega ^{ - 1/3} + 2.63\,\omega ^{1/3}L^{2}} \right]}\times
&& \nonumber\\ \times \frac{{d\omega} }{{\mathrm{e}^{\omega
/\left( {kT} \right)} - 1}}.&&
\end{eqnarray}

The expression in square brackets is a slowly varying function of
$\omega $. Taking into account that the main contribution to
\textit{W}$_{BBR}$ is due to frequencies near the ionization
threshold, one can replace $\omega $ by 1/(2\textit{n}$^{2}$).
After such replacement the integral in Eq.~(\ref{eq8}) can be
calculated analytically, and the final result is:
\begin{figure*}
\label{Fig2}
\includegraphics[width=18cm]{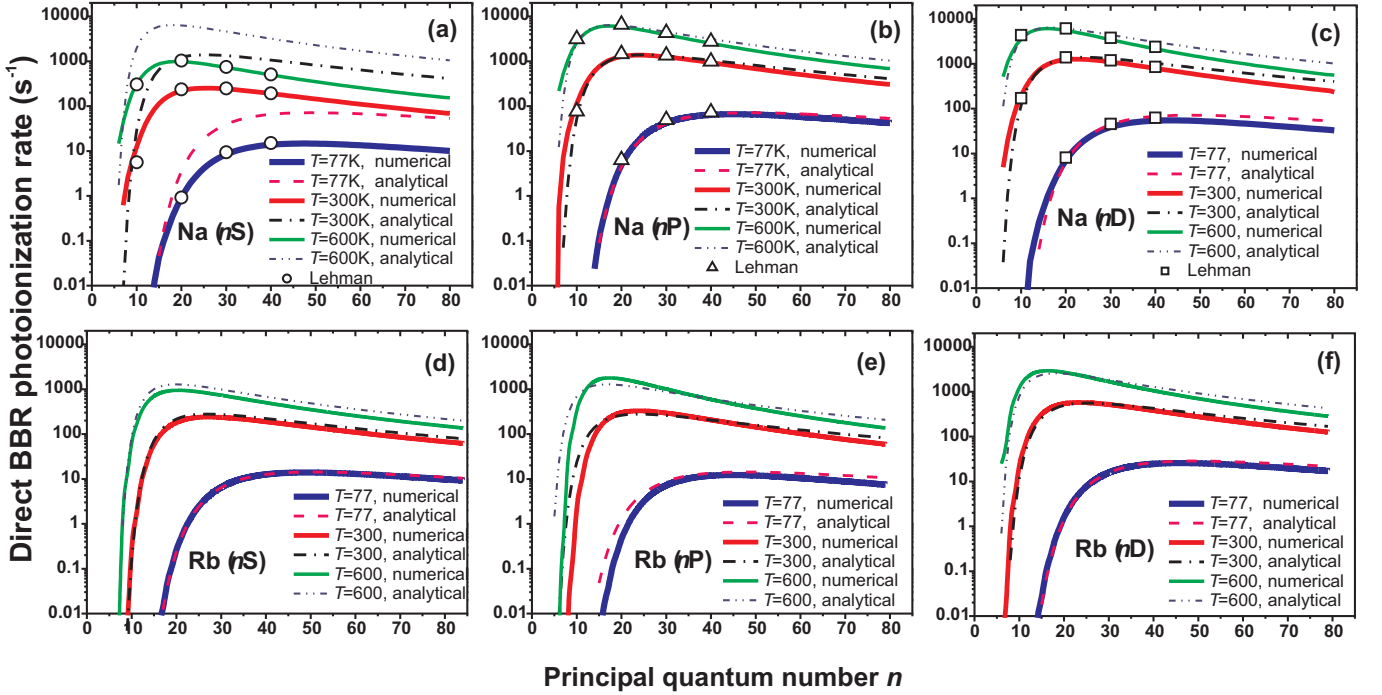}
\caption{Direct BBR-induced photoionization rate  of the sodium
\textit{n}S (a), \textit{n}P (b), \textit{n}D(c) and rubidium
\textit{n}S (d), \textit{n}P (e) and \textit{n}D (f) states with
\textit{n}=5-80 calculated at the ambient temperatures
\textit{T}=77, 300 and 600~K.}
\end{figure*}

\begin{equation}
\label{eq9}W_{BBR}\approx  \frac{{kT}}{{\pi ^{2}c^{3}}}\left[
{\frac{{2.80}}{{n^{7/3}}} + \frac{{2.09L^{2}}}{{n^{11/3}}}}
\right] \mathrm{ln}\left( {\frac{{1}}{{1 - \mathrm{exp}\left( { -
\frac{{\omega _{nL}} }{{kT}}} \right)}}} \right).
\end{equation}

Equation (\ref{eq9}) gives the approximate direct BBR
photoionization rate in atomic units for \textit{T} measured in
Kelvins. Alternatively, it can be rewritten to yield
\textit{W}$_{BBR}$ in the units of s$^{-1}$ for temperature
\textit{T} taken in Kelvins as follows:

\begin{eqnarray}
\label{eq10} W_{BBR} = C_{L} T\left[ {\frac{{14423}}{{n^{7/3}}} +
\frac{{10770L^{2}}}{{n^{11/3}}}} \right] \times \qquad \qquad
\qquad && \nonumber
\\ \times \mathrm{ln}\left( {\frac{{1}}{{1 - \mathrm{exp}\left( { -
\frac{{157890}}{{Tn^{2}}}} \right)}}} \right) \,\,\left[
{\mathrm{s}^{ - 1}} \right]. \quad &&
\end{eqnarray}

\noindent Here \textit{C}$_{L}$ is an \textit{L}-dependent scaling
coefficient, which will be discussed later. By replacing
\textit{n} with the effective principal quantum number,
Eq.~(\ref{eq10}) can be used for quick estimations of direct BBR
ionization rates for various Rydberg atoms and states. The precise
values, however, should be calculated numerically.

The results of our numerical and analytical calculations of the
direct BBR-induced photoionization rates for sodium and rubidium
\textit{n}S, \textit{n}P and \textit{n}D states with
\textit{n}=5-80 at the ambient temperatures \textit{T}=77, 300 and
600~K are shown in Fig.~2. A good agreement of our numerical
results obtained using the Dyachkov and Pankratov model with the
theoretical data obtained by Lehman~\cite{Lehman} is observed. For
the case of rubidium such comparison is not possible because no
other published data are available, to the best of our knowledge.

It is also interesting and instructive to compare the results
obtained by numeric calculation with those obtained using the
analytical formula (\ref{eq10}). Figure~2(a) shows a remarkable
disagreement between the numerical and analytical results
(assuming \textit{C}$_{L}$=1) for sodium \textit{n}S states, which
are known to possess a large quantum defect ($\delta _{S}$=1.348).
At the same time, Figs.~2(b) and 2(c) show that in the case of Na
\textit{n}P and \textit{n}D states, which have smaller quantum
defects ($\delta _{P}$=0.855 and $\delta _{D}$=0.015), the
agreement between the numerical and analytical results is much
better. This is not unexpected, since Eq.~(\ref{eq6}) is valid
only for states with small quantum defects. Formally, the
disagreement for the non-hydrogenic \textit{n}S states stems from
peculiarities of the asymptotic behavior of Bessel functions in
Eq.~(\ref{eq6}) for states with $L \ll 1$: the analytical
expression of GDK model yields close photoionization cross-section
values for \textit{n}S, \textit{n}P and \textit{n}D states, while
the accurate numerical calculations yield significantly smaller
cross-sections for sodium \textit{n}S states [see Fig.~2(a)]. At
the same time, one can see from Fig.~2(a) that shapes of the
analytical curves are quite similar to the numerical ones.
Therefore, one can simply introduce a scaling coefficient in
Eq.~(\ref{eq10}) in order to make it valid also for \textit{n}S
states.

In order to illustrate that, in Figs.~2(d)-(f) we show the
rescaled results of Eq.~(\ref{eq10}) for the case of Rb Rydberg
atoms \textit{n}S, \textit{n}P, and \textit{n}D states, which all
have large quantum defects ($\delta _{L}$ = 3.13, 2.66, and 1.34,
respectively). A good agreement between the analytical and
numerical results was obtained when the rate obtained from
Eq.~(\ref{eq10}) was scaled by a factor of \textit{C}$_{S}$=0.2,
\textit{C}$_{P}$=0.2 and \textit{C}$_{D}$=0.4 for the case of
\textit{n}S, \textit{n}P, and \textit{n}D states, respectively.

Our precise numerical data on \textit{W}$_{BBR}$ are summarized in
Tables (\ref{table1})-(\ref{table6}) of Appendix.


\subsection{BBR-induced mixing of Rydberg states}

BBR causes not only direct photoionization of the initially
populated levels. It also induces transitions between neighboring
Rydberg states, thus leading to a population redistribution
\cite{Galvez1, Galvez2, PaperII}. For example, after laser
excitation of the Na 16S state, the BBR-induced transitions
populate the neighboring $n'$P states [Fig.~1(a)]. The
calculations show that these states have significantly higher
direct photoionization rates \textit{W}$_{BBR}$ than the 16S state
itself. Hence, BBR-induced population transfer to $n'$P state can
noticeably affect the measured effective BBR ionization rate. The
rates of spontaneous and BBR-induced transitions from the initial
16S and 16D states to a number of $n'$P states have been
calculated by us in \cite{PaperI} and are shown in Figs.~1(b) and
1(c).

Importantly, absorption of BBR induces also transitions to higher
Rydberg states, which are denoted as $n''$ in Fig.~1(a). These
states can be field ionized by the electric field pulses usually
applied in experiments in order to extract ions into channeltron
or microchannel plate detectors.


\subsection{Field ionization of high Rydberg states populated by BBR}

Extraction electric field pulses, which are commonly used to
extract ions from the ionization zone to the ionization detector,
ionize Rydberg states with principal quantum numbers \textit{n}
exceeding some critical value $n_c$. This critical value $n_c$
depends on the amplitude of the applied electric field and it can
be found from the approximate formula \cite{SFI}

\begin{equation}
\label{eq11} E_{c} \approx 3.2 \cdot 10^{8}n_{c}^{ - 4} \quad
\left( \mathrm{V/cm} \right).
\end{equation}

\noindent were \textit{E}$_{c}$ is the critical electric field for
$n_{c}$. Hence, if a BBR mediated process populates a state with
$n' \ge n_c $, this state will be ionized and thus will contribute
to the detected ionization signal \cite{SpencerIon}.

We calculated the radial matrix elements $R\left( {nL \to n'L'}
\right)$ of all dipole-allowed transitions to other $n'L'$ states
with $L'=(L \pm 1)$ using the semi-classical formulae of
\cite{Dyachkov}. The rate of a BBR-induced transition between the
states $nL$ and $n'L'$ is expressed through the rate of
spontaneous transitions given by the Einstein coefficient $A\left(
{nL \to n'L'} \right)$:

\begin{eqnarray}
\label{eq12} A\left( {nL \to n'L'} \right) =
\frac{{4}}{{3c^{3}}}\frac{{L_{max}} }{{2L + 1}}R^{2}\left( {nL \to
n'L'} \right),&& \nonumber\\W\left( {nL \to n'L'} \right)=A\left(
{nL \to n'L'} \right)\frac{{\omega _{nn'}^{3}}
}{{\mathrm{e}^{\omega_{nn'}/\left({kT}\right)} - 1}},&&
\end{eqnarray}

\noindent where $\omega _{nn'} =1/(2n^2)-1/(2n'^2)$ is the
transition frequency. Note that spontaneous transitions are
possible only to lower levels, while BBR leads to transitions both
upwards and downwards.

The total rate $W_{SFI}$ of BBR transitions to all Rydberg states
with $n' \ge n_{c}$ was calculated by summing the individual
contributions of $nL \to n'L'$ transitions given by
Eq.~(\ref{eq12}):

\begin{equation}
\label{eq13} W_{SFI} = \,\sum\limits_{n' \ge n_{c}}
{\,\sum\limits_{L' = L \pm 1} {W\left( {nL \to n'L'} \right)}}.
\end{equation}

The values of $W_{SFI}$ were numerically calculated for various
amplitudes \textit{E} of the electric field pulses.

We also compared the numerical values with those obtained from the
approximate analytical formulae, which has been derived with the bound-bound
matrix elements of the GDK model:

\begin{eqnarray}
\label{eq14} W_{SFI}\approx \frac{{1}}{{\pi ^{2}c^{3}n^{3}}}\times
\qquad \qquad \qquad \qquad \qquad \qquad && \nonumber\\ \times
\int\limits_{1/(2n^2)-1/(2n_c^2)}^{\quad\omega_{nL}}  {\left[
{2.22\,\omega ^{ - 1/3} + 2.63\,\omega ^{1/3}L^{2}} \right]}\times
&& \nonumber\\  \times \frac{{d\omega} }{{\mathrm{e}^{\omega
/\left( {kT} \right)} - 1}}.\qquad &&
\end{eqnarray}

The integration limits are chosen such that the integral accounts
for transitions to those Rydberg states, for which
$\left[1/(2n^2)-1/(2n_c^2)\right]<\omega<\omega _{nL}$
 (i.e., states above the field
ionization threshold). Integration of Eq.~(\ref{eq14}) gives
another useful analytical formula that is similar to
Eq.~(\ref{eq10}):

\begin{eqnarray}
\label{eq15} W_{SFI} = C_{L} T\left[ {\frac{{14423}}{{n^{7/3}}} +
\frac{{10770L^{2}}}{{n^{11/3}}}} \right]\times \qquad \qquad
\qquad \qquad && \nonumber
\\ \times \left(\mathrm{ln}\frac{{1}}{{1 - \mathrm{exp}\left( {\frac{{157890}}{{Tn_{c}^{2}} } -
\frac{{157890}}{{Tn^{2}}}} \right)}}\,- \right. \qquad \qquad
&&\nonumber
\\- \left. \mathrm{ln}\frac{{1}}{{1 - \mathrm{exp}\left( { -
\frac{{157890}}{{Tn^{2}}}} \right)}} \right) \left[ \mathrm{s}^{ -
1}\right ],\qquad &&
\end{eqnarray}

\noindent
where \textit{T} is in Kelvins.

\begin{figure*}
\label{Fig3}
\includegraphics[width=18cm]{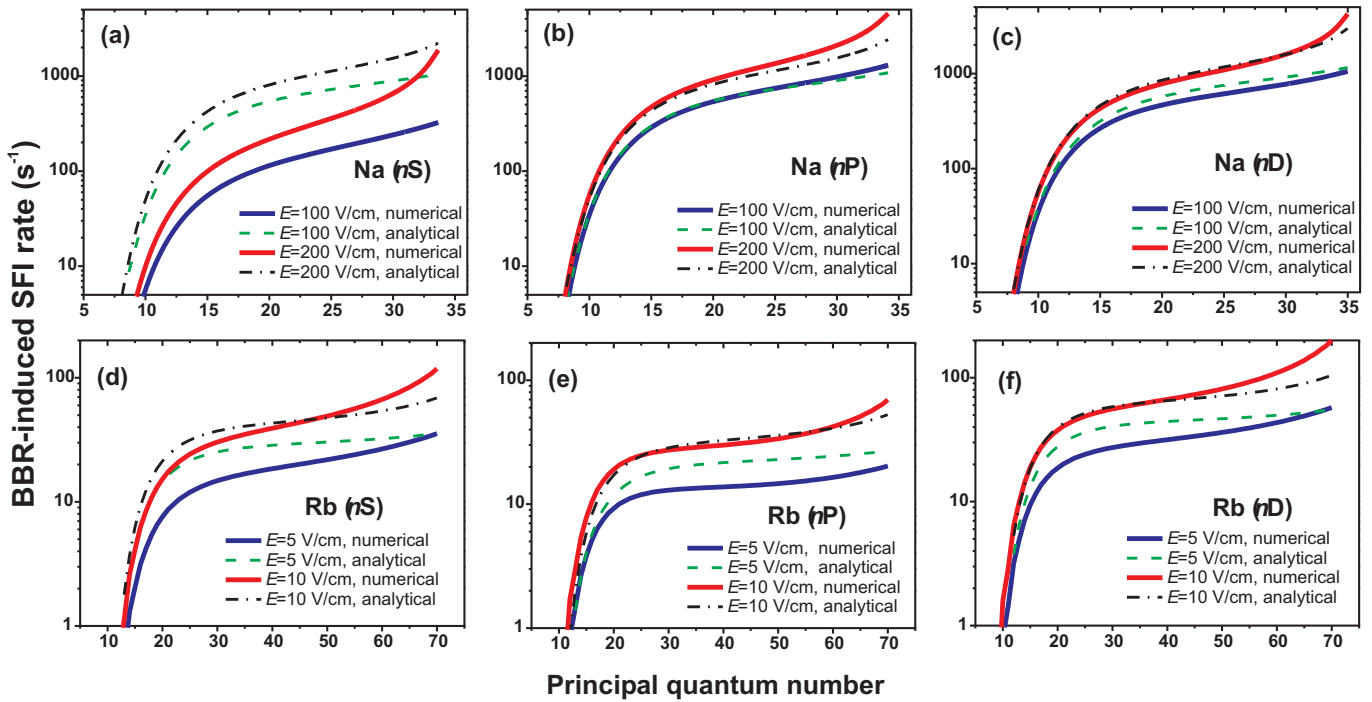}
\caption{The calculated BBR-induced SFI rate \textit{W}$_{SFI}$
for (a) Na(\textit{n}S), (b) Na(\textit{n}P), and (c)
Na(\textit{n}D) states with \textit{n}=5-35 at the electric field
amplitudes of \textit{E}=100 and 200~V/cm, and for the (d)
Rb(\textit{n}S), (e) Rb(\textit{n}P), and (f) Rb(\textit{n}D)
states with \textit{n}=5-80 at the electric field amplitudes of
\textit{E}=5 and 10~V/cm. In all cases the ambient temperature is
\textit{T}=300~K. }
\end{figure*}

The obtained numerical and analytical data on \textit{W}$_{SFI}$
are presented in Fig.~3. For Na atoms, the calculations were made
for the \textit{n}S, \textit{n}P and \textit{n}D states with
\textit{n}=5-35 at the ambient temperature \textit{T}=300~K
[Figs.~3(a), 3(b) and 3(c)]. The amplitudes \textit{E} of the
extracting electric field pulse are chosen as 100 and 200~V/cm
(corresponding to $n_{c} = 42$ and 36, respectively). These values
are close to those used in our recent atomic-beam experiment on
ionization of Na atoms \cite{PaperI}. Alternatively, Figs.~3(d),
3(e) and 3(f) show the calculated rates $W_{SFI}$ for \textit{n}S,
\textit{n}P and \textit{n}D states of Rb for the lower \textit{E}
values (5~V/cm and 10~V/cm corresponding to $n_{c}$=91 and $n_{c}$
=76, respectively), which are more adequate to experiments with
Rydberg atoms in cold gases.

In order to achieve a better agreement between the analytical
formula (\ref{eq15}) and the results of numerical calculations,
scaling coefficients $C_{L}$ in Eq.~(\ref{eq15}) were also
introduced. The best agreement was obtained at $C_{S}=0.2$,
$C_{P}=0.15$ and $C_{D}=0.3$ for the Rb \textit{n}S, \textit{n}P
and \textit{n}D states, respectively. In the case of \textit{n}P
and \textit{n}D states of Na data such scaling was not necessary,
while in the case of \textit{n}S states of Na the best agreement
[not shown in Fig.~3(a)] was found at $C_{S}=0.25$. The
possibility to achieve a satisfactory agreement between the
numerical and analytical data of Fig.~3 suggests that, using
appropriate scaling coefficients, the analytical formula
(\ref{eq15}) is also suitable to quick estimates of BBR-induced
SFI rates for various Rydberg atoms and states.

Our precise numerical data on $W_{SFI}$ are summarized in Tables
(\ref{table1})-(\ref{table6}) of Appendix.


\subsection{Ionization of Rydberg states populated by BBR}

In this section we shall analyze the influence of time evolution
of populations of Rydberg states upon interaction with ambient BBR
photons. The typical timing diagram of laser excitation of Rydberg
states and detection of ions is shown in Fig.~4. Such scheme was
used in our recent experiment on collisional ionization of Na
Rydberg atoms \cite{PaperI}. The first electric-field pulse is
applied immediately after the laser excitation pulse in order to
remove the atomic $A^{+}$ and molecular $A_{2}^{+}$ photoions
produced by the laser pulse. Then atoms are allowed to interact
with ambient BBR during the time interval (\textit{t}$_{1}$,
\textit{t}$_{2}$). The second electric-field pulse extracts the
ions, which have been produced by collisional and BBR-induced
ionization, to the ion detector. The atomic and molecular ions
were distinguished using a time-of-flight method. At the
ground-state density of $5\times 10^{10}$cm$^{-3}$  the BBR
ionization is the main source of atomic ions \cite{PaperI} and the
contribution from collisonal ionization of Penning type is
negligible.

Let us consider first the simplest case of laser excitation of a
single \textit{n}S state. The evolution of the number $N_{A^{ +} }
$ of atomic ions produced via absorption of BBR photons by atoms
in the initial \textit{n}S state is given by

\begin{equation}
\label{eq16}
\frac{{dN_{A^{ +} } \left( {t} \right)}}{{dt}} = W_{BBR} N_{nS} \left( {t}
\right),
\end{equation}

\noindent where
$N_{nS}(t)=N_{nS}(t=0)\mathrm{exp}(-t/\tau_{eff}^{nS})$ is the
total number of Rydberg atoms remaining in the nS state as a
function of time, and $\tau_{eff}^{nS}$ is an effective lifetime
of the $n$S state at a given ambient temperature. The registered
photoions are produced during the time interval ($t_1, t_2$). The
total number of ions can then be found by integrating
Eq.~(\ref{eq16}) from $t_1$ to $t_2$.

\begin{eqnarray}
\label{eq17} N_{A^{ +} } = N_{nS} \left( {t = 0}
\right)W_{BBR}\times \qquad \qquad \qquad &&\nonumber\\ \times
\tau _{eff}^{nS} \left[ {\mathrm{exp}\left( { - t_{1} /\tau
_{eff}^{nS}}  \right) - \mathrm{exp}\left( { - t_{2} /\tau
_{eff}^{nS}}  \right)} \right],&&
\end{eqnarray}

\noindent This result can be rewritten by introducing an effective
interaction time \cite{PaperI}:

\begin{eqnarray}
\label{eq18} N_{A^{ +}} = N_{nS}\left( {t = 0}\right)W_{BBR}
t_{eff}^{nS},\qquad \qquad \qquad \qquad &&\nonumber\\
t_{eff}^{nS} = \tau _{eff}^{nS} \left[ {\mathrm{exp}\left( { -
t_{1} /\tau _{eff}^{nS}}  \right) - \mathrm{exp}\left( { - t_{2}
/\tau _{eff}^{nS}}  \right)} \right].&&
\end{eqnarray}

Blackbody radiation also induces transitions to other Rydberg
states $n'$P, as was discussed in Section II~B. Evolution of
populations of these states is described by the rate equation

\begin{eqnarray}
\label{eq19}\frac{{dN_{{n}'P} \left( {t} \right)}}{{dt}} =
 \left[W\left( {nS \to {n}'P} \right)+A\left( {nS \to {n}'P} \right)\right]\,N_{nS} \left( {t}
\right)-&&\nonumber\\ -{N_{{n}'P} \left( {t}
\right)}/{\tau_{eff}^{n'P}},\qquad &&
\end{eqnarray}

\noindent where  $A(n\mathrm{S} \to n'\mathrm{P})$ and
$W(n\mathrm{S} \to n'\mathrm{P})$ are the rates of population of
the $n'$P state due to spontaneous transitions to lower levels and
BBR induced transitions upwards and downwards from the initial
\textit{n}S state, and $\tau_{eff}^{nP} $ is the effective
lifetime of the $n'$P state.

\begin{figure}
\label{Fig4}
\includegraphics[width=6cm]{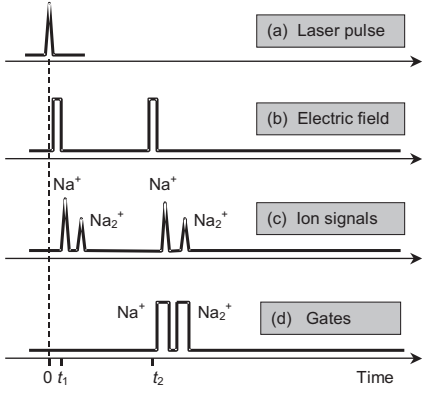}
\caption{Timing diagram of signals: (a) laser excitation pulse;
(b) electric field pulses for the ion extraction; (c) atomic $A^+$
and molecular $A^+_2$ ion signals; (d) detector gates. }
\end{figure}

A combination of Eq.~(\ref{eq19}) with Eqs.~(\ref{eq16}) and
(\ref{eq17}) yields

\begin{eqnarray}
\label{eq20} W_{BBR}^{mix} \left( {nS} \right) =
\sum\limits_{{n}'} {\frac{{\left[W\left( {nS \to {n}'P}
\right)+A\left( {nS \to {n}'P} \right)\right] }}{ \left(\tau
_{eff}^{n'P}\right)^{-1}- \left(\tau_{eff}^{nS}\right)^{-1} } }
\times &&\nonumber\\ \times W_{BBR} \left( {{n}'P} \right)\left(
{1 - \frac{{t_{eff}^{nS}} }{{t_{eff}^{{n}'P}} }} \right).\quad &&
\end{eqnarray}

The main contribution to the sum in Eq.~(\ref{eq20}) is from $n'$P
states with $n' =n \pm  1$ [see Fig.~1(b)]. The effective BBR
ionization rates for \textit{n}P and \textit{n}D states were
determined in the same way as for \textit{n}S states, taking into
account the population transfer to both $n'(L+1)$ and $n'(L-1)$
states.

The rate $W_{SFI}^{mix}$  describes the second-order process of
BBR-induced transitions from the neighboring $n'L'$ states to
highly excited states $n''L''$ with $n''>n_c$ [see Fig.~5(a)],
followed by ionization of these states by extracting electric
field pulses. This rate can be calculated using the same
Eq.~(\ref{eq20}), in which $W_{BBR}$  is replaced by $W_{SFI}$ and
the summation is done over the states with $n'<n_c$.

Our precise numerical data on $W_{BBR}^{mix}$ and $W_{SFI}^{mix}$
are summarized in Tables (\ref{table1})-(\ref{table6}) of
Appendix.




\section{Results of calculations and comparison with experiment}

\begin{figure}
\label{Fig5}
\includegraphics[width=6cm]{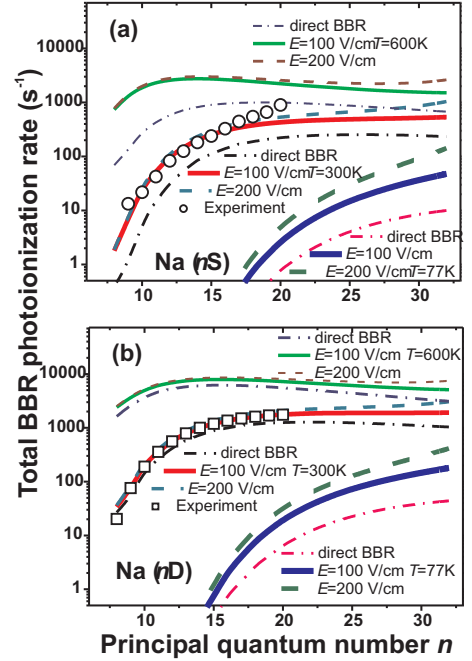}
\caption{Calculated total BBR-induced ionization rates
$W_{BBR}^{tot} $ for (a) \textit{n}S and (b) \textit{n}D states of
Na with \textit{n}=8-35 for the extracting electric field pulses
of 100~V/cm and 200~V/cm at the ambient temperatures of
\textit{T}=77, 300 and 600~K. Experimental points are taken from
Ref.~\cite{PaperI}. }
\end{figure}

\begin{figure*}
\label{Fig6}
\includegraphics[width=18cm]{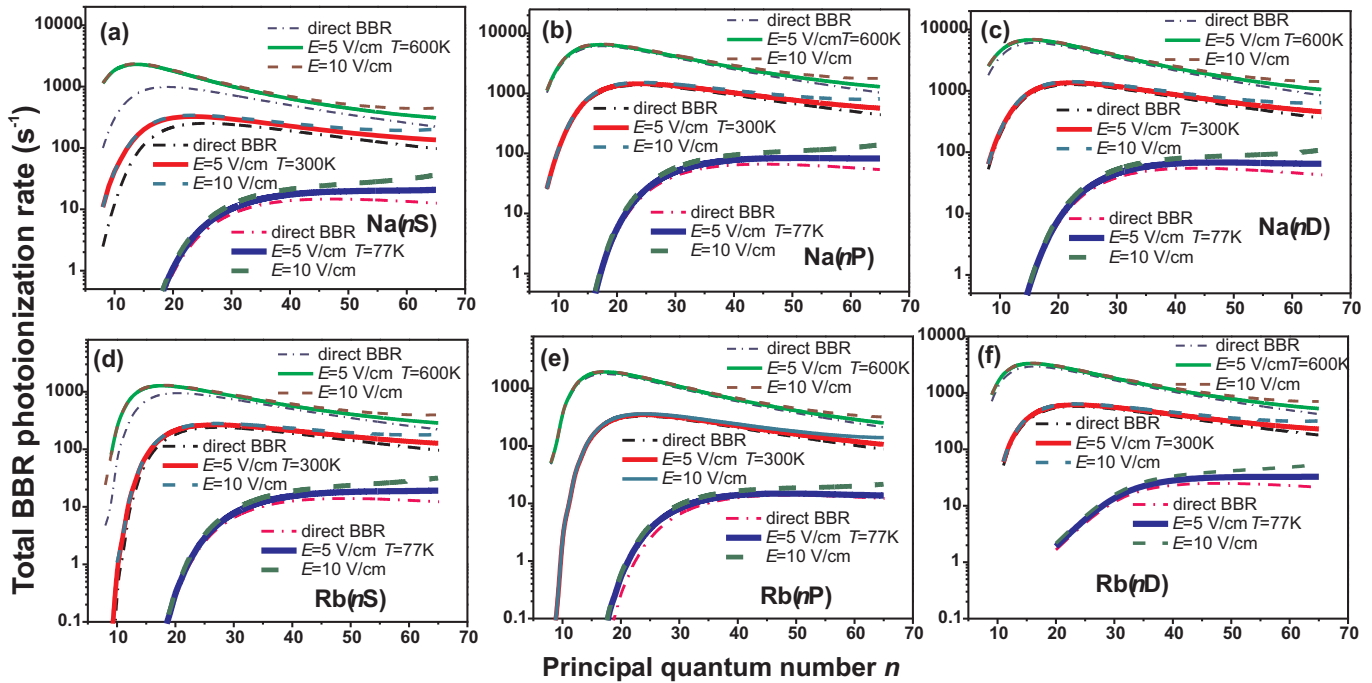}
\caption{The calculated total BBR-induced ionization rates
$W_{BBR}^{tot} $ for (a) Na(\textit{n}S), (b) Na(\textit{n}P), (c)
Na(\textit{n}D), (d) Rb(\textit{n}S), (e) Rb(\textit{n}P), and (d)
Rb(\textit{n}D) Rydberg states for the electric field amplitudes
of 5~V/cm and 10~V/cm at the ambient temperatures of
\textit{T}=77~K, 300~K and 600~K. }
\end{figure*}

Calculated total BBR-induced ionization rates $W_{BBR}^{tot} $ for
Na \textit{n}S, \textit{n}P and \textit{n}D states at the
temperatures \textit{T} = 77, 300, 600~K and for the extracting
electric field pulses of \textit{E}=100~V/cm and 200~V/cm, as well
as their comparison with our experimental data from
Ref.~\cite{PaperI}, are shown in Fig.~5. Since the values of
$W_{BBR}^{tot}$ depend on the timing of the experiment (see
Fig.~4), the calculations were performed for $t_1=0.3 \, \mu$s and
$t_2=2.1 \, \mu$s, which were used in our experiment
\cite{PaperI}. At the 300~K temperature a good agreement is
observed for \textit{n}D states with \textit{n}=8-20 and for
\textit{n}S states with \textit{n}=8-15. At the same time, for
\textit{n}S states with $n>15$ the calculated ionization rates
start to deviate from the experimental values; the measured values
exceed the theoretical ones by a factor of 2.1 for \textit{n}=20,
and the shape of the experimental \textit{n}-dependence differs
from the theoretical one.

One possible explanation of such anomaly for \textit{n}S states is
related to their specific orbit that penetrates into the atomic
core. The penetration causes a strong interaction between the
Rydberg electron and the core, e.g., due to core polarization
\cite{Aymar}. This results in large quantum defect and a Cooper
minimum in the photoionization cross-sections. This assumption is
supported by the good agreement of theory and experiment for the
hydrogen-like \textit{n}D states, which have a small quantum
defect and almost non-penetrating orbits.

Total BBR-induced ionization rates were also calculated for Na and
Rb $n$S, $n$P and $n$D states in a broader range of $n$ and for
lower amplitudes of the electric field pulses (5~V/cm and
10~V/cm). Such fields are more relevant to the experiments with
cold Rydberg atoms, e.g., to ultracold plasma formation from cold
Rydberg atoms clouds \cite{Robinson, Li}, since these experiments
explore the atoms excited to relatively high Rydberg states that
are ionized in weaker electric fields (35~V/cm for $n\sim 50$ and
10~V/cm for $n\sim 80$). The calculated ionization rates are
presented in Fig.~6. It can be seen that SFI and BBR-induced
level-mixing processes alter the shapes of \textit{n}-dependences
of total ionization rates $W_{BBR}^{tot} $, which effect is more
pronounced at lower temperatures and larger \textit{n}. As Fig.~6
has a logarithmic scale, the visible effect of additional
processes looks small. However, from the tables of Appendix it can
be seen that, e.g., for the Rb~65S state at $T$=300~K the total
BBR ionization rate is twice larger than the direct
photoionization rate (see Table~5).

Precise results of our numerical calculations for a five-step set
of principal quantum numbers \textit{n} are summarized in Tables
(\ref{table1})-(\ref{table6}) of Appendix. The values of
ionization rates at the other \textit{n} values can be obtained
either by extrapolation or upon request from the authors of this
paper.


\section{Conclusion}

We have calculated the total BBR-induced ionization rates of Na
and Rb \textit{n}S, \textit{n}P and \textit{n}D Rydberg states for
principal quantum numbers  \textit{n}=8-65 at the ambient
temperatures of 77, 300 and 600~K. Our calculations take into
account the effect of BBR-induced mixing of Rydberg states and
their field ionization by extracting electric field pulses. Useful
analytical formulae have been derived, which allow for quick
estimation of ionization rates and their dependences on the
principal quantum number \textit{n}. The numerical results are in
a good agreement with our recent experiment data on Na \textit{n}S
and \textit{n}D states, except for \textit{n}S states with $n>15$,
which is most probably associated with the Cooper minimum in the
photoionization cross-section.

The obtained results show that BBR-induced redistribution of
population over Rydberg states and their field ionization by
extracting electric fields affect both the magnitudes of total
ionization rates and shapes of their dependencies on the principal
quantum number. This suggests that these processes are important
and cannot be ignored in the calculations and measurements of BBR
ionization rates. Equations (\ref{eq16})-(\ref{eq20}), as well as
the analytical formulae (\ref{eq9}) and (\ref{eq15}), can be used
to calculate total ionization rates $W_{BBR}^{tot} $ under
particular experimental conditions. The obtained numerical results
may be helpful to the analysis of ionization signals measured in
experiments on collisional ionization and spontaneous formation of
ultracold plasma, since BBR-induced ionization is the main channel
of delivering atomic ions. At the same time, as we have revealed
that theoretical data for Na $n$S-states noticeably disagree with
experiment at $15 <n \leq 20$, new experimental data for Na and Rb
in a broader range of principal quantum numbers would be of
interest for the further improvement of theory, especially for the
non-hydrogen-like states.


\begin{acknowledgments}
This work was supported by INTAS grant No. 04-83-3692, Russian
Foundation for Basic Research (grants No. 05-02-16181,
05-03-33252), Siberian Branch of RAS, EU FP6 TOK project LAMOL,
European Social Fund, Latvian Science Council, and NATO grant
EAP.RIG.981387.
\end{acknowledgments}




\appendix*




\begin{table*}
\caption{\label{table1}Calculated BBR-induced ionization rates
(s$^{-1}$) for Na at \textit{T}=77~K}

\begin{tabular*}{\textwidth}{@{\extracolsep{\fill}}|c|c|c|c|c|c|c|c|c|}  \hline
\multicolumn{9}{|c|}{\textbf{\textit{n}S}}
\\\hline
& & & & & & & &\\
$n$&$W_{BBR}$&$W_{SFI}$&$W_{SFI}$&$W_{BBR}^{mix}$&$W_{SFI}^{mix}$&$W_{SFI}^{mix}$&$W_{BBR}^{tot}$&$W_{BBR}^{tot}$\\
&direct& 100V/cm & 200V/cm & &100V/cm& 200V/cm &100V/cm &200V/cm
\\\hline
& & & & & & & &\\
10& $ 1.56\times 10^{-7}$& $3.44 \times
10^{-9}$& $7.71 \times 10^{-9}$& $4.92 \times 10^{-18}$& $3.55
\times 10^{-9}$&$7.03\times 10^{-9}$& $2.12\times
10^{-7}$&$2.2\times 10^{-7}$
\\\hline
15& 0.009& 0.019& 0.042& 0.002& 0.003& 0.006& 0.033&0.059
\\\hline
20& 0.805& 1.655& 3.83& 0.059& 0.118& 0.240& 2.64&4.94
\\\hline
25& 4.04& 10.23& 25.29& 0.185& 0.420& 0.891& 14.9&30.4
\\\hline
30& 8.45& 27.12& 81.09& 0.257& 0.706& 1.689& 36.5&91.5
\\\hline
& & & & & & & &\\
&$W_{BBR}$&$W_{SFI}$&$W_{SFI}$&$W_{BBR}^{mix}$&$W_{SFI}^{mix}$&$W_{SFI}^{mix}$&$W_{BBR}^{tot}$&$W_{BBR}^{tot}$\\
&direct& 5V/cm & 10V/cm & &5V/cm& 10V/cm &5V/cm &10V/cm
\\\hline
35& 11.8& 2.47& 5.20& 0.267& 0.0488& 0.107& 14.6&17.4
\\\hline
45& 14.6& 4.12& 9.09& 0.195& 0.0466& 0.106& 19.0&24.0
\\\hline
55& 14.1& 5.76& 13.9& 0.124& 0.0412& 0.102& 20.1&28.3
\\\hline
65& 12.6& 8.01& 23.6& 0.0773& 0.0377& 0.111& 20.7&36.4
\\\hline
\multicolumn{9}{|c|}{\textbf{\textit{n}P}}
\\\hline
& & & & & & & &\\
$n$&$W_{BBR}$&$W_{SFI}$&$W_{SFI}$&$W_{BBR}^{mix}$&$W_{SFI}^{mix}$&$W_{SFI}^{mix}$&$W_{BBR}^{tot}$&$W_{BBR}^{tot}$\\
&direct& 100V/cm & 200V/cm & &100V/cm& 200V/cm &100V/cm &200V/cm
\\\hline
& & & & & & & &\\
10& $7.09\times 10^{-6}$& $3.49 \times 10^{-7}$&
$6.91 \times 10^{-7}$& $4.75 \times 10^{-8}$& $1.51\times
10^{-8}$& $2.99\times 10^{-8}$& $7.5\times 10^{-6}$&$7.85 \times
10^{-6}$
\\\hline
15& 0.0996& 0.195& 0.389& 0.00111& 0.00200& 0.00409& 0.298& 0.493
\\\hline
20& 5.22& 10.2& 20.5& 0.0217& 0.0455& 0.0977& 15.4&25.8
\\\hline
25& 22.1& 50.8& 107& 0.0545& 0.132& 0.308& 73.1&130
\\\hline
30& 42.4& 120& 290& 0.0662& 0.197& 0.554& 163&333
\\\hline
& & & & & & & &\\
&$W_{BBR}$&$W_{SFI}$&$W_{SFI}$&$W_{BBR}^{mix}$&$W_{SFI}^{mix}$&$W_{SFI}^{mix}$&$W_{BBR}^{tot}$&$W_{BBR}^{tot}$\\
&direct& 5V/cm & 10V/cm & &5V/cm& 10V/cm &5V/cm &10V/cm
\\\hline
35& 57.0& 10.5& 22.9& 0.0726& 0.0144& 0.0318& 67.6& 80.0
\\\hline
45& 66.2& 16.2& 37.1& 0.0514& 0.0135& 0.0313& 82.5&103
\\\hline
55& 61.9& 21.6& 53.7& 0.0320& 0.0117& 0.0297& 83.5&116
\\\hline
65& 53.8& 28.6& 86.5& 0.0186& 0.00997& 0.0304& 82.4&140
\\\hline
\multicolumn{9}{|c|}{\textbf{\textit{n}D}}
\\\hline
& & & & & & & &\\
$n$&$W_{BBR}$&$W_{SFI}$&$W_{SFI}$&$W_{BBR}^{mix}$&$W_{SFI}^{mix}$&$W_{SFI}^{mix}$&$W_{BBR}^{tot}$&$W_{BBR}^{tot}$\\
&direct& 100V/cm & 200V/cm & &100V/cm& 200V/cm &100V/cm &200V/cm
\\\hline
& & & & & & & &\\
10& $1.11 \times 10^{-4}$& $1.53\times 10^{-5}$&
$2.98 \times 10^{-5}$& $7.61 \times 10^{-7}$& $4.96 \times
10^{-7}$& $9.82\times 10^{-7}$&$1.28\times 10^{-4}$ & $1.43\times
10^{-4}$
\\\hline
15& 0.251& 0.483& 0.947& 0.00270& 0.00472& 0.00953&0.742&1.21
\\\hline
20& 6.59& 12.7& 25.2& 0.0371& 0.0762& 0.158&19.4& 32.0
\\\hline
25& 22.6& 50.0& 104& 0.0926& 0.215& 0.467&72.9& 127
\\\hline
30& 39.0& 107& 256& 0.120& 0.331& 0.812&146& 296
\\\hline
& & & & & & & &\\
&$W_{BBR}$&$W_{SFI}$&$W_{SFI}$&$W_{BBR}^{mix}$&$W_{SFI}^{mix}$&$W_{SFI}^{mix}$&$W_{BBR}^{tot}$&$W_{BBR}^{tot}$\\
&direct& 5V/cm & 10V/cm & &5V/cm& 10V/cm &5V/cm &10V/cm
\\\hline
35& 49.3& 8.88& 19.4& 0.131& 0.0239& 0.0527&58.3& 68.8
\\\hline
45& 54.6& 13.0& 29.6& 0.0995& 0.0230& 0.0528&67.7&84.3
\\\hline
55& 49.8& 16.7& 41.5& 0.0663& 0.0204& 0.0503&66.7& 91.5
\\\hline
65& 42.7& 21.8& 65.9& 0.0424& 0.0180& 0.0519&64.6&109
\\\hline

\end{tabular*}

\end{table*}




\begin{table*}
\caption{\label{table2}Calculated BBR-induced ionization rates
(s$^{-1}$) for Na at \textit{T}=300~K}

\begin{tabular*}{\textwidth}{@{\extracolsep{\fill}}|c|c|c|c|c|c|c|c|c|}  \hline
\multicolumn{9}{|c|}{\textbf{\textit{n}S}}
\\\hline
& & & & & & & &\\
$n$&$W_{BBR}$&$W_{SFI}$&$W_{SFI}$&$W_{BBR}^{mix}$&$W_{SFI}^{mix}$&$W_{SFI}^{mix}$&$W_{BBR}^{tot}$&$W_{BBR}^{tot}$\\
&direct& 100V/cm & 200V/cm & &100V/cm& 200V/cm &100V/cm &200V/cm
\\\hline
& & & & & & & &\\
10& 7.50& 1.33& 2.34& 13.6& 3.02& 4.88&25.5&28.3
\\\hline
15& 114& 39.4& 70.2& 78.6& 24.2& 39.6& 256& 302
\\ \hline
20& 223& 99.2& 185& 77.6& 29.2& 49.2& 429& 534
\\ \hline
25& 254& 155& 314& 53.6& 26.2& 46.9& 489& 669
\\ \hline
30& 242& 218& 537& 34.2& 23.2& 46.9& 518& 861
\\ \hline & & & & & & & &\\
&$W_{BBR}$&$W_{SFI}$&$W_{SFI}$&$W_{BBR}^{mix}$&$W_{SFI}^{mix}$&$W_{SFI}^{mix}$&$W_{BBR}^{tot}$&$W_{BBR}^{tot}$\\
&direct& 5V/cm & 10V/cm & &5V/cm& 10V/cm &5V/cm &10V/cm
\\\hline
35& 217& 19.3& 39.5& 22.6& 1.55& 3.29& 261& 283
\\\hline
45&166&22.8& 49.1& 10.2& 1.04& 2.30& 200& 227
\\\hline
55& 126&27.3& 64.2& 5.15& 0.780& 1.87& 159& 197
\\\hline
65& 97.4&35.0& 100& 2.80& 0.652& 1.86& 136& 202
\\\hline
\multicolumn{9}{|c|}{\textbf{\textit{n}P}}\\\hline & & & & & & &
&\\
$n$&$W_{BBR}$&$W_{SFI}$&$W_{SFI}$&$W_{BBR}^{mix}$&$W_{SFI}^{mix}$&$W_{SFI}^{mix}$&$W_{BBR}^{tot}$&$W_{BBR}^{tot}$\\
&direct& 100V/cm & 200V/cm & &100V/cm& 200V/cm &100V/cm &200V/cm
\\\hline
10& 79.92& 15.99& 25.84& 2.426& 0.8949& 1.456& 99.23& 109.6
\\\hline
15& 778.2& 240.9& 393.6& 17.82& 5.476& 9.174&1042& 1199
\\\hline
20& 1295& 501.8& 845.4& 19.05& 6.384& 11.24& 1823& 2171
\\\hline
25& 1374& 709.0& 1272& 13.50& 5.610& 10.76& 2102& 2670
\\\hline
30& 1253& 932.0& 1919& 8.444& 4.660& 10.56& 2198& 3191
\\\hline & & & & & & & &\\
&$W_{BBR}$&$W_{SFI}$&$W_{SFI}$&$W_{BBR}^{mix}$&$W_{SFI}^{mix}$&$W_{SFI}^{mix}$&$W_{BBR}^{tot}$&$W_{BBR}^{tot}$\\
&direct& 5V/cm & 10V/cm & &5V/cm& 10V/cm &5V/cm &10V/cm
\\\hline
35& 1090& 79.97& 169.6& 6.145& 0.3428& 0.7388& 1177& 1267
\\\hline
45& 795.2& 89.28& 198.3& 2.830& 0.2290& 0.5203& 887.5&996.9
\\\hline
55& 586.0& 102.1& 247.0& 1.419& 0.1679&0.4172& 689.7& 834.8
\\\hline
65& 441.8& 124.9& 366.0& 0.7254& 0.1267& 0.3768&567.5& 808.9
\\\hline
\multicolumn{9}{|c|}{\textbf{\textit{n}D}}\\\hline & & & & & & &
&\\
$n$&$W_{BBR}$&$W_{SFI}$&$W_{SFI}$&$W_{BBR}^{mix}$&$W_{SFI}^{mix}$&$W_{SFI}^{mix}$&$W_{BBR}^{tot}$&$W_{BBR}^{tot}$\\
&direct& 100V/cm & 200V/cm & &100V/cm& 200V/cm &100V/cm &200V/cm
\\\hline
10& 157.5& 37.01& 59.22& 13.71& 2.596& 4.237& 210.8& 234.7
\\\hline
15& 890.1& 268.2& 434.4& 39.52& 8.003& 13.30&1206& 1377
\\\hline
20& 1253& 467.7& 782.0& 36.76& 8.176& 14.05& 1765& 2086
\\\hline
25& 1244& 613.5& 1094& 26.17& 6.907& 12.61& 1891& 2377
\\\hline
30& 1099& 774.0& 1592& 17.38& 5.679& 11.70& 1896& 2720
\\\hline & & & & & & & &\\
&$W_{BBR}$&$W_{SFI}$&$W_{SFI}$&$W_{BBR}^{mix}$&$W_{SFI}^{mix}$&$W_{SFI}^{mix}$&$W_{BBR}^{tot}$&$W_{BBR}^{tot}$\\
&direct& 5V/cm & 10V/cm & &5V/cm& 10V/cm &5V/cm &10V/cm
\\\hline
35& 936.2& 65.00& 137.7& 12.31& 0.4207& 0.8980& 1014& 1087
\\\hline
45& 666.5& 70.18& 155.6& 6.040& 0.2783& 0.6197&743.0& 828.8
\\\hline
55& 483.1& 78.52& 189.6& 3.240& 0.2025& 0.4856&565.1& 676.4
\\\hline
65& 360.0& 94.60& 277.3& 1.849& 0.1557& 0.4332&456.6& 639.6
\\\hline

\end{tabular*}

\end{table*}




\begin{table*}
\caption{\label{table3}Calculated BBR-induced ionization rates
(s$^{-1}$) for Na at \textit{T}=600 K}

\begin{tabular*}{\textwidth}{@{\extracolsep{\fill}}|c|c|c|c|c|c|c|c|c|}  \hline
\multicolumn{9}{|c|}{\textbf{\textit{n}S}}
\\\hline & & & & & & & &\\
$n$&$W_{BBR}$&$W_{SFI}$&$W_{SFI}$&$W_{BBR}^{mix}$&$W_{SFI}^{mix}$&$W_{SFI}^{mix}$&$W_{BBR}^{tot}$&$W_{BBR}^{tot}$\\
&direct& 100V/cm & 200V/cm & &100V/cm& 200V/cm &100V/cm &200V/cm
\\\hline
10& 284.4& 41.90& 71.08& 1396& 187.6& 294.9& 1910& 2046 \\ \hline
15& 910.1& 185.6& 321.5& 1396& 237.4& 378.9& 2729& 3007\\ \hline
20& 991.5& 290.5& 526.8& 770.7& 175.8& 289.9& 2229& 2579\\ \hline
25& 874.7& 378.8& 748.8& 411.1& 130.9& 229.0& 1796& 2264\\ \hline
30& 728.5& 489.6& 1172& 229.0& 105.5& 208.4& 1553& 2338\\ \hline &
& & & & & & &\\
&$W_{BBR}$&$W_{SFI}$&$W_{SFI}$&$W_{BBR}^{mix}$&$W_{SFI}^{mix}$&$W_{SFI}^{mix}$&$W_{BBR}^{tot}$&$W_{BBR}^{tot}$\\
&direct& 5V/cm & 10V/cm & &5V/cm& 10V/cm &5V/cm &10V/cm
\\\hline
35& 601& 42.9& 87.6& 139& 7.04& 14.8& 790& 843 \\ \hline
45&416&48.3& 104& 57.1& 4.46& 9.82& 526& 586 \\ \hline
55&299&56.5& 132& 27.5& 3.25& 7.76& 386& 467 \\ \hline
65&222&71.3& 203& 14.6& 2.67& 7.57& 311& 448 \\ \hline
\multicolumn{9}{|c|}{\textbf{\textit{n}P}}
\\\hline & & & & & & & &\\
$n$&$W_{BBR}$&$W_{SFI}$&$W_{SFI}$&$W_{BBR}^{mix}$&$W_{SFI}^{mix}$&$W_{SFI}^{mix}$&$W_{BBR}^{tot}$&$W_{BBR}^{tot}$\\
&direct& 100V/cm & 200V/cm & &100V/cm& 200V/cm &100V/cm &200V/cm
\\\hline
10& 2580& 353.0& 554.1& 190.1& 31.47& 49.95& 3155& 3374 \\\hline
15& 6054& 1054& 1681& 279.2& 45.01& 73.60& 7433& 8088\\\hline
20&5900& 1433& 2364& 173.8& 34.97& 60.14& 7541& 8497 \\\hline
25&4884& 1715& 3015& 96.59& 26.01& 48.63& 6722& 8044 \\ \hline
30&3898& 2077& 4183& 52.98& 19.78& 43.40& 6048& 8178 \\ \hline

& & & & & & & &\\
&$W_{BBR}$&$W_{SFI}$&$W_{SFI}$&$W_{BBR}^{mix}$&$W_{SFI}^{mix}$&$W_{SFI}^{mix}$&$W_{BBR}^{tot}$&$W_{BBR}^{tot}$\\
&direct& 5V/cm & 10V/cm & &5V/cm& 10V/cm &5V/cm &10V/cm
\\\hline
35& 3115& 177.6& 375.1& 35.25& 1.458& 3.135& 3330& 3529 \\ \hline
45& 2049& 188.9& 417.8& 14.72& 0.9238& 2.094& 2254& 2484\\ \hline
55& 1426& 210.8& 508.0& 7.008& 0.6582& 1.632& 1644& 1942\\ \hline
65& 1029& 254.5& 742.4& 3.470& 0.4856& 1.439& 1288& 1777\\ \hline
35& 3115& 177.6& 375.1& 35.25& 1.458& 3.135& 3330& 3529\\ \hline

\multicolumn{9}{|c|}{\textbf{\textit{n}D}}
\\\hline & & & & & & & &\\
$n$&$W_{BBR}$&$W_{SFI}$&$W_{SFI}$&$W_{BBR}^{mix}$&$W_{SFI}^{mix}$&$W_{SFI}^{mix}$&$W_{BBR}^{tot}$&$W_{BBR}^{tot}$\\
&direct& 100V/cm & 200V/cm & &100V/cm& 200V/cm &100V/cm &200V/cm
\\\hline
10& 3679& 506.3& 787.9& 750.7& 68.35& 108.5& 5004& 5326 \\ \hline
15& 6226& 1052& 1666& 631.2& 58.47& 94.75& 7968& 8618 \\ \hline
20& 5547& 1286& 2106& 368.8& 40.12& 67.39& 7242& 8090 \\ \hline
25& 4405& 1458& 2549& 211.4& 28.43& 50.80& 6103& 7216 \\ \hline
30& 3434& 1709& 3438& 124.9& 21.11& 42.54& 5288& 7039 \\ \hline

& & & & & & & &\\
&$W_{BBR}$&$W_{SFI}$&$W_{SFI}$&$W_{BBR}^{mix}$&$W_{SFI}^{mix}$&$W_{SFI}^{mix}$&$W_{BBR}^{tot}$&$W_{BBR}^{tot}$\\
&direct& 5V/cm & 10V/cm & &5V/cm& 10V/cm &5V/cm &10V/cm
\\\hline
35& 2702& 143.5& 302.9& 81.66& 1.594& 3.390& 2929& 3090 \\ \hline
45& 1733& 148.1& 327.0& 36.94& 0.9921& 2.202& 1919& 2099\\ \hline
55& 1185& 162.0& 389.4& 19.08& 0.6971& 1.667& 1367& 1595\\ \hline
65& 844.1& 192.7& 562.0& 10.72& 0.5188& 1.438& 1048& 1418\\ \hline
\end{tabular*}

\end{table*}




\begin{table*}
\caption{\label{table4}Calculated BBR-induced ionization rates
(s$^{-1}$) for Rb at \textit{T}=77 K}

\begin{tabular*}{\textwidth}{@{\extracolsep{\fill}}|c|c|c|c|c|c|c|c|c|}  \hline
\multicolumn{9}{|c|}{\textbf{\textit{n}S}}
\\\hline
& & & & & & & &\\
$n$&$W_{BBR}$&$W_{SFI}$&$W_{SFI}$&$W_{BBR}^{mix}$&$W_{SFI}^{mix}$&$W_{SFI}^{mix}$&$W_{BBR}^{tot}$&$W_{BBR}^{tot}$\\
&direct& 5V/cm & 10V/cm & &5V/cm& 10V/cm &5V/cm &10V/cm
\\\hline
& & & & & & & &\\
10& $3.34\times 10^{-12}$& $4.54\times
10^{-13}$& $9.46 \times 10^{-13}$& $3.36\times 10^{-13}$&
$4.36\times 10^{-14}$& $9.21\times 10^{-14}$& $4.18\times
10^{-12}$ & $4.72\times 10^{-12}$
\\ \hline
20& 0.2826& 0.04174& 0.08730& 0.005807& $7.58\times 10^{-4}$&
0.001622& 0.3309 & 0.3773
\\\hline
30& 6.541& 1.108& 2.343& 0.04932& 0.006986& 0.01519&7.706 & 8.949
\\ \hline
40& 12.61& 2.740& 5.967& 0.04750& 0.008213& 0.01840&15.41 & 18.64
\\ \hline
50& 13.91& 4.201& 9.678& 0.03120& 0.007135& 0.01687&18.15 & 23.64
\\ \hline
60& 12.97& 5.813& 14.96& 0.01917& 0.006156& 0.01615&18.81 & 27.97
\\ \hline
65& 12.21& 6.863& 19.53& 0.01502& 0.005856& 0.01688&19.10 & 31.78
\\ \hline
\multicolumn{9}{|c|}{\textbf{\textit{n}P}}
\\\hline
& & & & & & & &\\
10& $2.96\times 10^{-10}$& $3.84\times
10^{-11}$& $8.1\times 10^{-11}$& $2.51\times 10^{-13}$&
$3.39\times 10^{-14}$& $7.1\times 10^{-14}$& $3.35\times 10^{-10}$
& $3.77\times 10^{-10}$
\\\hline
20& 0.489& 0.0636& 0.136& 0.00443& $6.31 \times 10^{-4}$& 0.00132&
0.558 & 0.631
\\ \hline
30& 7.21& 1.02& 2.22& 0.0412& 0.00656& 0.0139&8.28 & 9.49
\\\hline
40& 11.8& 2.08& 4.65& 0.0447& 0.00887& 0.0193& 14.0 & 16.5
\\\hline
50& 11.9& 2.83& 6.70& 0.0321& 0.00854& 0.0195&14.8 & 18.7
\\\hline
60&10.5& 3.60& 9.49& 0.0212& 0.00797& 0.0201&14.1 & 20.0
\\\hline
65& 9.65& 4.09& 11.9& 0.0171& 0.00785& 0.0217&13.8 & 21.6
\\\hline
\multicolumn{9}{|c|}{\textbf{\textit{n}D}}
\\\hline
& & & & & & & &\\
 10& $5.11\times 10^{-7}$& $6.76\times 10^{-8}$&
$1.42\times 10^{-7}$& $4.23\times 10^{-9}$& $5.57\times 10^{-10}$&
$1.18\times 10^{-9}$& $5.83\times 10^{-7}$& $6.58\times 10^{-7}$
\\\hline
20& 1.66& 0.227& 0.475& 0.0149& 0.00207& 0.00442 & 1.90& 2.15
\\\hline
30& 15.7& 2.46& 5.19& 0.0710& 0.0110&0.0237 & 18.2& 21.0
\\\hline
40& 24.7& 4.99& 10.8& 0.0662& 0.0127&0.0280 & 29.7& 35.6
\\\hline
50& 25.4& 7.15& 16.5& 0.0450& 0.0113& 0.0264 & 32.6& 41.9
\\\hline
60& 22.9& 9.60& 24.9& 0.0287& 0.0101&0.0260 & 32.5& 47.8
\\ \hline
65& 21.3& 11.3& 32.5& 0.0229& 0.00969&0.0277 & 32.6& 53.9
\\\hline
\end{tabular*}

\end{table*}




\begin{table*}
\caption{\label{table5}Calculated BBR-induced ionization rates
(s$^{-1}$) for Rb at \textit{T}=300 K}

\begin{tabular*}{\textwidth}{@{\extracolsep{\fill}}|c|c|c|c|c|c|c|c|c|}  \hline
\multicolumn{9}{|c|}{\textbf{\textit{n}S}}
\\\hline
& & & & & & & &\\
$n$&$W_{BBR}$&$W_{SFI}$&$W_{SFI}$&$W_{BBR}^{mix}$&$W_{SFI}^{mix}$&$W_{SFI}^{mix}$&$W_{BBR}^{tot}$&$W_{BBR}^{tot}$\\
&direct& 5V/cm & 10V/cm & &5V/cm& 10V/cm &5V/cm &10V/cm
\\\hline
10& 0.604& 0.0195& 0.0390& 0.408& 0.0113& 0.0228&1.04&1.07
\\\hline
20& 183& 7.58& 15.3& 20.8& 0.633& 1.30& 212& 220\\\hline
30&235&14.8& 30.3& 9.25& 0.390& 0.819& 260& 276 \\ \hline
40&189&18.5& 39.2& 3.70& 0.227& 0.492& 211& 232\\ \hline
50&143&21.9& 49.1& 1.67& 0.150& 0.344& 167& 194 \\ \hline
60&109&26.7& 67.0& 0.842& 0.113& 0.288& 137& 177\\ \hline
65&96.1&30.3& 83.9& 0.619& 0.103& 0.287& 127& 181 \\ \hline
\multicolumn{9}{|c|}{\textbf{\textit{n}P}}
\\\hline
10& 3.69& 0.102& 0.206& 0.0598& 0.00432& 0.00873& 3.85& 3.96 \\
\hline 20& 300& 9.25& 19.0& 8.14& 0.311& 0.626& 317& 327
\\ \hline
30& 294& 13.0& 27.3& 4.87& 0.255& 0.520& 312& 327 \\ \hline
40&207& 13.8& 30.0& 2.20& 0.173& 0.363& 223& 240 \\ \hline
50&144&14.7& 33.8& 1.07& 0.126& 0.278& 160& 179 \\ \hline
60&103&16.5& 42.3& 0.568& 0.103& 0.246& 120& 146\\ \hline
65&87.8&18.0& 51.0& 0.426& 0.0965& 0.251& 106& 140 \\ \hline
\multicolumn{9}{|c|}{\textbf{\textit{n}D}}
\\\hline
10& 27.8& 0.782& 1.58& 2.73& 0.0532& 0.108& 31.4& 32.2 \\ \hline
20& 540& 18.7& 37.6& 21.4& 0.626& 1.28& 581& 600 \\ \hline
30&517&27.3& 55.9& 10.4& 0.439& 0.908& 555& 584\\ \hline
40&381&31.6& 66.9& 4.60& 0.282& 0.600& 418& 453 \\ \hline
50&277&36.2& 81.2& 2.25& 0.199& 0.444& 315& 360 \\ \hline
60&205&43.5& 110& 1.20& 0.157& 0.385& 250& 316\\ \hline
65&179&49.1& 138& 0.904& 0.146& 0.391& 229& 318 \\ \hline

\end{tabular*}

\end{table*}




\begin{table*}
\caption{\label{table6}Calculated BBR-induced ionization rates
(s$^{-1}$) for Rb at \textit{T}=600 K}

\begin{tabular*}{\textwidth}{@{\extracolsep{\fill}}|c|c|c|c|c|c|c|c|c|}  \hline
\multicolumn{9}{|c|}{\textbf{\textit{n}S}}
\\\hline & & & & & & & &\\
$n$&$W_{BBR}$&$W_{SFI}$&$W_{SFI}$&$W_{BBR}^{mix}$&$W_{SFI}^{mix}$&$W_{SFI}^{mix}$&$W_{BBR}^{tot}$&$W_{BBR}^{tot}$\\
&direct& 5V/cm & 10V/cm & &5V/cm& 10V/cm &5V/cm &10V/cm
\\\hline
10& 81.20& 1.429& 2.838& 177.4& 2.395 & 4.806 & 262.4&266.2
\\\hline
20& 941.6& 26.22& 52.51& 259.5& 4.569 & 9.340 & 1232& 1263
\\\hline
30& 734.1& 35.55& 72.58& 68.99& 1.925 & 4.022 & 840.6& 879.7
\\\hline
40& 500.8& 40.35& 85.19& 23.30& 1.009 & 2.181 & 565.4& 611.4
\\\hline
50& 350.5& 45.92& 102.7& 9.740&0.6394 &1.458 & 406.8&464.4
\\ \hline
60& 255.3& 54.93& 137.1& 4.738&0.4709 &1.188 & 315.4&398.3
\\\hline
65& 220.8& 61.92& 170.6& 3.445&0.4253 &1.176 & 286.6&396.0
\\\hline

\multicolumn{9}{|c|}{\textbf{\textit{n}P}}
\\\hline

10& 373.9& 5.011& 10.06& 20.19 & 0.6466& 1.299 & 399.7& 405.5
\\\hline
20& 1690& 30.95& 63.33& 100.7 & 2.073&4.156 & 1824&1858
\\\hline
30& 1009& 30.95& 64.80& 35.35 & 1.170&2.375 & 1077&1112
\\\hline
40& 592.4& 30.00& 64.98&13.44 & 0.7148&1.488 & 636.5&672.3
\\\hline
50& 374.1& 30.71& 70.43& 6.029 & 0.4986& 1.086 & 411.3& 451.6
\\\hline
60& 252.3& 33.85& 86.50&3.076 & 0.3940&0.9362 & 289.6& 342.8
\\\hline 65& 211.3& 36.78& 103.7&2.279 & 0.3675&0.9453 & 250.7&318.2
\\\hline
\multicolumn{9}{|c|}{\textbf{\textit{n}D}}
\\\hline
10& 1263& 17.45& 34.98& 279.7 & 2.919 & 5.900&1563 & 1584
\\\hline
20& 2696& 57.56& 115.5& 220.0 & 3.948 & 8.007&2978 & 3040
\\\hline
30& 1681& 63.97& 130.5&71.48&2.059 & 4.235&1819 & 1887
\\ \hline
40& 1051& 68.29& 144.0& 27.59 & 1.211 & 2.556&1148 & 1225
\\\hline
50& 699.6& 75.56& 168.9&12.59 & 0.8201 & 1.814& 788.6 & 883.0
\\\hline 60& 492.7& 89.13& 223.8& 6.510 & 0.6320 & 1.534& 589.0 &
724.5 \\\hline 65& 420.5& 100.1& 280.1& 4.847 &0.5827 &
1.544&526.1 & 706.9 \\ \hline

\end{tabular*}

\end{table*}




\clearpage
\begin{thebibliography}{10}

\bibitem{Gallagher}T.~F.~Gallagher and W.~E.~Cooke, Phys. Rev. Lett.
\textbf{42}, 835-839 (1979).

\bibitem{SpencerIon}W.~P.~Spencer, A.~G.~Vaidyanathan, D.~Kleppner, and
T.~W.~Ducas, Phys. Rev. A 26, 1490-1493 (1982).

\bibitem{Beiting}G.~F.~Hildebrandt, E.~J.~Beiting, C.~Higgs,
G.~J.~Hatton, K.~A.~Smith, F.~B.~Dunning, and R.~F.~Stebbings,
Phys. Rev. A 23, 2978-2982 (1981).

\bibitem{Spencer}W.~P.~Spencer, A.~G.~Vaidyanathan, D.~Kleppner, and
T.~W.~Ducas, Phys. Rev. A \textbf{25}, 380-384 (1982).

\bibitem{Rydberg atoms}T.~F.~Gallagher, \textit{Rydberg Atoms}
(Cambridge: Cambridge University Press) (1994).

\bibitem{Farley}J.~W.~Farley and W.~H.~Wing, Phys. Rev. A \textbf{23}, 2397-2424
(1981).

\bibitem{Galvez1} E.~J.~Galvez, J.~R.~Lewis, B.~Chaudhuri, J.~J.~Rasweiler,
H.~Latvakoski, F.~De~Zela, E.~Massoni, and H.~Castillo, Phys. Rev.
A \textbf{51}, 4010-4017 (1995).

\bibitem{Galvez2} E.~J.~Galvez, C.~W.~MacGregor, B.~Chaudhuri, S.~Gupta,
E.~Massoni and F.~De~Zela, Phys. Rev. A \textbf{55}, 3002-3006
(1997).


\bibitem{Robinson} M.~P.~Robinson, B.~Laburthe~Tolra, Michael~W.~Noel,
T.~F.~Gallagher and P.~Pillet, Phys. Rev. Let. \textbf{85,} 4466
(2000).

\bibitem{Li} W.~Li, M.~W.~Noel, M.~P.~Robinson, P.~J.~Tanner,
T.~F.~Gallagher, D.~Comparat, B.~Laburthe~Tolra, N.~Vanhaecke,
T.~Vogt, N.~Zahzam, P.~Pillet, D.~A.~Tate, Phys. Rev. A
\textbf{70}, 042713 (2004).

\bibitem{PaperI}
I.~I.~Ryabtsev, D.~B.~Tretyakov, I.~I.~Beterov, N.~N.~Bezuglov,
K.~Miculis and A.~Ekers, J. Phys. B \textbf{38}, S17 (2005).

\bibitem{Lehman}G.~W.~Lehman, J.\textbf{} Phys. B: At. Mol. Phys.
\textbf{16} 2145-2156 (1983).


\bibitem{Dyachkov}L.~G.~Dyachkov and P.~M.~Pankratov, J. Phys. B \textbf{27},
461 (1994).

\bibitem{GDK} S.~P.~Goreslavsky, N.~B.~Delone and V.~P.~Krainov, Sov. Phys.
JETP, \textbf{55} 246 (1982)

\bibitem{Davydkin}V.~A.~Davydkin and B.~A.~Zon, Opt. Spectr. \textbf{51}
13 (1981).

\bibitem{Aymar} M.~Aymar, J. Phys. B \textbf{11}, 1413 (1978).

\bibitem{SFI}
R.~F.~Stebbings, C.~J.~Latimer, W.~P.~West, F.~B.~Dunning, and
T.~B.~Cook, Phys. Rev. A \textbf{12}, 1453 (1975); T.~Ducas,
M.~G.~Littman, R.~R.~Freeman, and D.~Kleppner, Phys. Rev. Lett.
\textbf{35}, 366 (1975); T.~F.~Gallagher, L.~M.~Humphrey,
R.~M.~Hill, and S.~A.~Edelstein, Phys. Rev. Lett. \textbf{37},
1465 (1976).

\bibitem{PaperII} K.~Miculis, I.~I.~Beterov, N.~N.~Bezuglov, I.~I.~Ryabtsev, D.~B.~Tretyakov,
A.~Ekers and A.~N.~Klucharev, Phys.B: At. Mol. Opt. Phys.
\textbf{38} 1811 (2005).

\bibitem{PaperIII} I.~I.~Beterov, D.~B.~Tretyakov, I.~I.~Ryabtsev, N.~N.~Bezuglov, K.~Miculis,
A.~Ekers and A.~N.~Klucharev, J. Phys. B: At. Mol. Opt. Phys.
\textbf{38} 4349 (2005).

\end{thebibliography}
\end{document}